\documentstyle[sprocl]{article}
\input{psfig}
\def\be{\begin{equation}}
\def\ee{\end{equation}}
\def\bea{\begin{eqnarray}}
\def\eea{\end{eqnarray}}
\def\lsim{\raise0.3ex\hbox{$\;<$\kern-0.75em\raise-1.1ex\hbox{$\sim\;$}}}
\def\gsim{\raise0.3ex\hbox{$\;>$\kern-0.75em\raise-1.1ex\hbox{$\sim\;$}}}

\begin{document}
\hfill{\bf CERN-TH/95-285}

\title{$\nu_\tau$ OSCILLATION EXPERIMENTS AND PRESENT DATA \cite{ours}}

\author{M.\ C.\ Gonzalez-Garcia\footnote{Talk Given at the {\sl International
Workshop on Elementary Particle Physics, Present and Future}, Valencia, Spain
5--9 June 1995}}

\address{Theory Division,   CERN,
CH-1211 Geneva 23, Switzerland.}

\maketitle\abstracts{Our goal in this paper is to examine the discovery
potential of laboratory
experiments searching for the oscillation $\nu_\mu(\nu_e) \rightarrow
\nu_\tau$, in the light of recent data on solar and atmospheric neutrino
experiments, which we analyse together with the most restrictive results from
laboratory experiments on neutrino oscillations in a four-neutrino framework.}

\noindent
{\bf CERN-TH/95-285} \\
\noindent
{\bf October 1995}
\section{Introduction}
If neutrinos have a mass, a neutrino produced with flavour $\alpha$,
after travelling a distance $L$, can be  detected in the
charged-current (CC) reaction $\nu \; N' \rightarrow  l_\beta \; N $ with a
probability
\begin{equation}
\begin{array}{ll}
\langle P_{\alpha \beta}\rangle &=
 {\displaystyle \delta_{\alpha,\beta}-4\sum_{i=1}^n\sum_{j=i+1}^n
\mbox{Re}[ U_{\alpha i}U^\star_{\beta i} U^\star_{\alpha j} U_{\beta j}]
\langle\sin^2\left(\frac{\Delta_{ij}}{2}\right)\rangle },
\end{array}
\end{equation}
where $U$ is the mixing matrix. The average includes the dependence on  the
neutrino energy spectrum, the cross section for the process in which the
neutrino is detected, and detection efficiency for the experiment. The
probability, therefore, oscillates with oscillation lengths
$\frac{\Delta_{ij}}{2}=1.27 \frac{|m_i^2-m_j^2|}{\mbox{eV}^2}
\frac{L/E}{\mbox{m/MeV}}$ where $E$ is the neutrino energy. For oscillation
lengths such that $\Delta m^2_{ij} \gg 1/(L/E)$ the oscillating phase will have
been over many cycles before the detection and therefore  it will have averaged
to ${1}/{2}$. On the other hand, for $\Delta m^2_{ij} \ll 1/(L/E)$, the
oscillation did not  have time to give any effect.

Present data from solar and atmospheric neutrino experiments favour the
hypothesis of neutrino oscillations. All  solar neutrino experiments
\cite{solar} find less $\nu_e$ than predicted theoretically.  As for
atmospheric neutrino experiments, two of them \cite{kamisub,kamimul,IMB}
measure a ratio  $\nu_\mu/\nu_e$ smaller than expected from theoretical
calculations. Nevertheless, this interpretation  requires confirmation from
further experiments, in particular from laboratory experiments, where the
experimental conditions, in particular the shape, energy, and flux of the
neutrino beam are under control.

At present all laboratory neutrino experiments report no evidence for
neutrino oscillation \cite{laboratory} with the possible  exception of LSND
\cite{LSND}, which looks for $\nu_\mu \rightarrow \nu_e$ oscillations.
In addition a number of new experiments are in the process of starting
taking data of being propossed both at CERN \cite{CERN} and Fermilab
\cite{FNAL}.
Our goal is to examine the discovery potential of these
experiments searching for the oscillation $\nu_\mu(\nu_e) \rightarrow
\nu_\tau$, in the light of recent data on solar and atmospheric neutrino
experiments, which we analyse together with the most restrictive results from
laboratory experiments on neutrino oscillations in a four-neutrino
framework \cite{ours}.

\section{Four-Flavour Models}
Naive two-family counting shows that it is very difficult to  fit all
experimental information mentioned above with three neutrino flavours.  In the
spirit of Pauli, one is tempted to introduce a new neutrino as a ``desperate
solution'' to  understand all present data. The nature of such a particle is
constrained by  LEP results on the invisible $Z$ width as well as data on the
primordial $^4$He abundance. Those rule out the existence of additional,
light, active neutrinos. In consequence the fourth neutrino state must be
sterile.

If one assumes a {\sl natural} mass hierarchy  with two
light neutrinos with their main projection in the $\nu_s$ and $\nu_e$
directions and two heavy neutrinos with their largest component along the
$\nu_\mu$ and $\nu_\tau$ flavours the mixing matrix
can be parametrized in a general way as
\begin{equation}
U=\begin{array}{c|cccc}
 & \nu_s & \nu_e & \nu_\mu & \nu_\tau \\
\hline
\nu_1 & c_{es} & s_{es}c_{e\mu}c_{e\tau} & s_{es}s_{e\mu}
& s_{es}c_{e\mu}s_{e\tau} \\
\nu_2 & -s_{es} & c_{es}c_{e\mu}c_{e\tau} & c_{es} s_{e\mu}
& c_{es}c_{e\mu} s_{e\tau} \\
\nu_3 & 0 & -c_{\mu\tau}s_{e\mu}c_{e\tau}-s_{\mu\tau}s_{e\tau}
& c_{\mu\tau}c_{e\mu} & -c_{\mu\tau}s_{e\mu}s_{e\tau}+s_{\mu\tau}c_{e\tau} \\
\nu_4 & 0 & s_{\mu\tau}s_{e\mu}c_{e\tau}-c_{\mu\tau}s_{e\tau}
& -s_{\mu\tau}c_{e\mu} & s_{\mu\tau}s_{e\mu}s_{e\tau}+c_{\mu\tau}c_{e\tau} \\
\end{array}
\end{equation}
with $c_i=\cos\theta_i$ and $s_i=\sin\theta_i$.  For the sake of simplicity we
have assumed no CP violation in the lepton sector.  We also required that the
sterile neutrino does not mix directly with the two heavy states  to verify the
constraints from big bang  nucleosynthesis \cite{BBN}. Such a  hierarchy
appears naturally, for instance if one advocates an $L_e\pm L_\mu\mp L_\tau$
discrete symmetry for the mass matrix \cite{valle1}. In Ref.\ \cite{moha} a
similar mass pattern is also generated via a combination of see-saw mechanism
and loop mechanism. In this approximation  $m_1,m_2\ll m_3,m_4$ and
$|m_3^2-m_4^2| \ll m_3^2,m_4^2$.

The value of $m_3\approx m_4$ is inferred from the dark matter data.
Currently, the best scenario to explain the data considers a mixture of 70$\%$
cold plus 30$\%$ hot dark matter \cite{dark1}.  This implies $m_3\simeq
m_4=2$--$3.5$ eV. Such a mass pattern has been argued to yield satisfactory
results in Cold+Hot Dark Matter scenarios \cite{dark2}.
We define
\begin{equation}
\begin{array}{lr}
\Delta m^2_{solar}= |m_1^2-m_2^2| \;\;\;\; \Delta m^2_{AT}=|m_3^2-m_4^2| \\
\Delta M^2_{DM}=|m_1^2-m_3^2|\simeq |m_1^2-m_4^2|\simeq |m_2^2-m_3^2|
\simeq |m_2^2-m_4^2|\simeq (4-10)\; \mbox{eV}^2\;. &\\
\label{our}
\end{array}
\end{equation}
Transition probabilities between the different flavours will  have  therefore
contributions from the three oscillation lengths due to the three different
mass differences in the problem which we will denote
$\sin^2({\Delta_{solar}}/{2})$,  $\sin^2({\Delta_{AT}}/{2})$, and
$\sin^2({\Delta_{DM}}/{2})$, respectively.

\section{Global Analysis}
\label{global}
At present the most precise laboratory experiments searching for neutrino
oscillations are \cite{laboratory} the reactor experiment at Bugey, which looks
for
$\nu_e$  disappearance, and the CDHSW experiment  at CERN,  which
searches for $\nu_\mu$ disappearance.  The E776 experiment at BNL
searchs for the $\bar\nu_\mu\rightarrow \bar\nu_e$ appearance channel and the
E531 experiment at Fermilab  for the  $\nu_\mu\rightarrow \nu_\tau$
channel. Neither of these experiments shows evidence for neutrino oscillation
on those channels.  Recently the Liquid Scintillator Neutrino Detector (LSND)
experiment \cite{LSND} has announced the observation of an anomaly that can be
interpreted as  neutrino oscillations in the channel $\bar{\nu_\mu} \rightarrow
\bar{\nu_e}$. Most of the oscillation parameters required as explanation are
already excluded by the E776  and KARMEN  experiments.

For the Bugey reactor experiment the relevant transition probability is the
$\nu_e$ survival probability. For any value of the atmospheric mass difference
this probability will always verify
\begin{equation}
0.93 \lsim P_{ee}^{Bugey} \leq  1-
2 c_{e\mu}^2 c_{e\tau}^2(1- c_{e\mu}^2 c_{e\tau}^2)
\;\; \Rightarrow \;\;  c_{e\mu}^2 c_{e\tau}^2\geq 0.96\; .
\label{bugey}
\end{equation}
For CDHSW the relevant probability is the $\nu_\mu$ survival probability
\begin{equation}
0.95\lsim P_{\mu\mu}^{CDHSW}\leq 1- \frac{1}{2}\sin^2(2 \theta_{e\mu})
\;\; \Rightarrow \;\;  \sin^2(2 \theta_{e\mu})\lsim 0.1\; .
\end{equation}
For E776 the situation is somehow more involved,  since the value of the
oscillating phase $\langle \sin^2( {\Delta_{DM}}/{2})\rangle $ varies
in the range $\Delta_{DM}=4$--$10$ eV$^2$ due to the wiggles of the resolution
function of the
experiment. Also, the experiment is sensitive to the
atmospheric mass difference. We find that the limit
$1.5\times 10^{-3}\geq  P_{e\mu}^{E776}$  is verified
for any value of the atmospheric mass difference and the $\mu\tau$ mixing
angle, if
\begin{equation}
\sin^2(2\theta_{e\mu}) c^2_{e\tau}\leq (2\mbox{--}5)\times 10^{-3}
\end{equation}
The limit from E531 on the mixings $e\mu$ and $e\tau$  is always less
restrictive than the previous ones for any value of $\Delta m^2_{AT}$ and
$\Delta M^2_{DM}$.

Combining these constraints we obtain that $e\mu$ and $e\tau$ mixings are
constrained to
\begin{equation}
\begin{array}{lr}
\sin^2 (2\theta_{e\mu}) \leq (2\mbox{--}5)\times 10^{-3} &\;\;\;\;\;\;
\sin^2 (2\theta_{e\tau}) \leq 0.16\;\;\;\;
\end{array}
\label{acc}
\end{equation}
where the range of $\sin^2 (2\theta_{e\mu})$ depends on the specific value
of $\Delta M^2_{DM}$.

If we now turn to the effect due to the oscillation with $\Delta_{AT}$,
we can rewrite the relevant probabilities for the different experiments
expanding in the small angles $e\mu$ and $e\tau$:
\begin{equation}
\begin{array}{l}
P_{\mu\mu}^{CDHSW}\simeq
1-\frac{1}{2}\sin^2(2\theta_{e\mu})-\sin^2(2\theta_{\mu\tau})
\sin^2(\frac{\Delta_{AT}}{2})\\
P_{\mu\tau}^{E531}\simeq \sin^2(2\theta_{\mu\tau}) c^2_{e\tau}\sin^2(\frac
{\Delta_{AT}}{2})\\
P_{e\mu}^{776}\simeq \sin^2(2\theta_{e\mu})c^2_{e\tau}
\sin^2(\frac{\Delta_{DM}}{2})
+ \sin^2(2\theta_{\mu\tau}) s^2_{e\tau} \sin^2(\frac{\Delta_{AT}}{2}) \; .\\
\end{array}
\end{equation}
With the constraints in Eq.\ (\ref{acc}), the Bugey experiment is not sensitive
to oscillations with $\Delta_{AT}$.
The relevant exclusion contours for each channel are shown in the Figures.

We now turn to the atmospheric neutrino data.
Neutrinos are produced when cosmic rays
hit the atmosphere and initiate atmospheric cascades. The mesons
present in the cascade decay leading to a flux of $\nu_e$ and $\nu_\mu$ which
reach the Earth and interact in the different neutrino detectors. Naively the
expected ratio of $\nu_\mu$ to $\nu_e$ is in the proportion $2:1$, since the
main reaction is $\pi \rightarrow \mu \nu_\mu$ followed by $\mu \rightarrow e
\nu_\mu \nu_e$. However, the expected ratio of muon-like interactions to
electron-like interactions in each experiment depends on the detector
thresholds and efficiencies  as well as on the expected neutrino fluxes.
Currently four experiments have observed atmospheric neutrino interactions.
Two experiments, Kamiokande
\cite{kamisub,kamimul} and IMB \cite{IMB}, have observed  a ratio
of $\nu_\mu$-induced events to $\nu_e$-induced events  smaller than the
expected one. In particular Kamiokande has performed two different analyses for
both sub-GeV neutrinos \cite{kamisub} and multi-GeV neutrinos \cite{kamimul},
which show the same deficit. On the other hand, the results from
 Fr\'ejus  and NUSEX
\cite{frejus} appear to be in agreement with the predictions.

The results of the three most precise experiments  in terms of
the double ratio  $R_{\mu/e}/R^{MC}_{\mu/e}$
of experimental-to-expected ratio of
muon-like to electron-like events are
\begin{equation}
\begin{array}{l}
R_{\mu/e}/R^{MC}_{\mu/e}=0.55 \pm 0.11 \;\;\;\;\mbox{for IMB}\\
R_{\mu/e}/R^{MC}_{\mu/e}=0.6 \pm 0.09 \;\;\;\;\mbox{for Kamiokande sub-GeV}\\
R_{\mu/e}/R^{MC}_{\mu/e}=0.59 \pm 0.1 \;\;\;\;\mbox{for Kamiokande
multi-GeV}\\R_{\mu/e}/R^{MC}_{\mu/e}=1.06 \pm 0.23 \;\;\;\;\mbox{for Fr\'ejus}
\end{array}
\label{atmosdat}
\end{equation}
The statistical and systematic errors have been added in quadrature.  The
systematic error contains a 5 \% contribution due to the neutrino flux
uncertainties. For the Montecarlo prediction we  have used the expected fluxes
from \cite{gaisser}  depending
on the neutrino energies. Use of other flux calculations would yield similar
numbers.

In each experiment the number of $\mu$ events, $N_\mu$, and of $e$ events,
$N_e$, in the presence of oscillations will be
\begin{equation}
N_\mu=N^0_{\mu\mu} \langle P_{\mu\mu}\rangle +N^0_{e\mu} \
\langle P_{e\mu}\rangle \; ,  \;\;\;\;\;\
N_e=N^0_{ee} \langle P_{ee}\rangle +N^0_{\mu e} \langle P_{\mu e}\rangle \; ,
\end{equation}
where
\begin{equation}
N^0_{\alpha\beta}=\int \frac{d^2\Phi_\alpha}{dE_\nu d\cos\theta_\nu}
\frac{d\sigma}{dE_\beta}\epsilon(E_\beta)
dE_\nu dE_\beta d(\cos\theta_\nu)
\end{equation}
and
\begin{equation}
\langle P_{\alpha\beta}\rangle =\frac{1}{N^0_{\alpha\beta}}\int
\frac{d^2\Phi_\alpha}{dE_\nu d\cos\theta_\nu} P_{\alpha\beta}
\frac{d\sigma}{dE_\beta}\epsilon(E_\beta)
dE_\nu dE_\beta d(\cos\theta_\nu)\; .
\end{equation}
Here $E_\nu$ is the neutrino energy and $\Phi_\alpha$ is the flux of
atmospheric
neutrinos $\nu_\alpha$; $E_\beta$ is the final charged lepton energy and
$\epsilon(E_\beta)$ is the detection efficiency for such charged lepton;
$\sigma$ is the interaction cross section $\nu \; N \rightarrow N'\; l$.
The expected rate with no oscillation would be $
R^{MC}_{\mu/e}= {N^0_{\mu\mu}}/{N^0_{ee}}$ .
The double ratio $R_{\mu/e}/R^{MC}_{\mu/e}$  is then given by
\begin{equation}
\frac{R_{\mu/e}}{R^{MC}_{\mu/e}}=
\frac{\langle P_{\mu\mu}\rangle +\frac{N^0_{e\mu}}{N^0_{\mu\mu}}
\langle P_{e\mu}\rangle }
{\langle P_{ee}\rangle +\frac{N^0_{\mu e}}{N^0_{ee}}
\langle P_{\mu e}\rangle  } \; .
\label{atratio}
\end{equation}
We perform a global fit to the data in Eq.\ (\ref{atmosdat}).
In Fig.\ \ref{mutauat1} the results are shown for zero mixings $e\mu$ and
$e\tau$ as
in a two-family scenario. Figure  \ref{mutauat2} shows the effect of the
inclusion of
the mixings. As seen in the figure the inclusion of the $e\tau$ mixing leads
to a  more constrained area for the oscillation parameters. The effect of
the $e\tau$ mixing  is to increase the value of the double
ratio since there is a decrease on the number of $\nu_e$. Therefore a larger
amount of $\mu\tau$ oscillation is needed to account
for the deficit. Due to the small values allowed, a non-zero mixing $e\mu$ does
not modify the  analysis of the atmospheric neutrino data.

Finally just to coment that for solar neutrino experiments the presence of new
mixings affect very little the analysis performed in the  two family scenario,
the small mixing MSW \cite{msw} solution is still valid \cite{sterile}.
The large mixing
solution is also in conflict with the constraints from  big bang
nucleosynthesis  \cite{BBN}.

\section {$\nu_\tau$ Oscillation Experiments:Discovery Potential}
\label{experiments}
\begin{table}
\caption{Summary of the performance parameters for CHORUS and NOMAD.}
\label{tab:performance}
\begin{tabular}{|c|c|c|c|c|c|c|}
\hline
 & NOMAD & CHORUS & E803/ & MINOS \\
Sensitivity: & &  & NAUSICAA (CERN) &  \\
$P_{e\tau}$ & $1.3 \times 10^{-2}$ & $0.8 \times 10^{-2}$
& $7 \times 10^{-4}$ & --      \\
$P_{\mu\tau}$ & $2\times 10^{-4}$ & $1.4\times 10^{-4}$ &
$1.4 \times 10^{-5}$ & 0.012   \\
\hline
\end{tabular}
\end{table}
\begin{center}
\begin{figure}
\psfig{figure=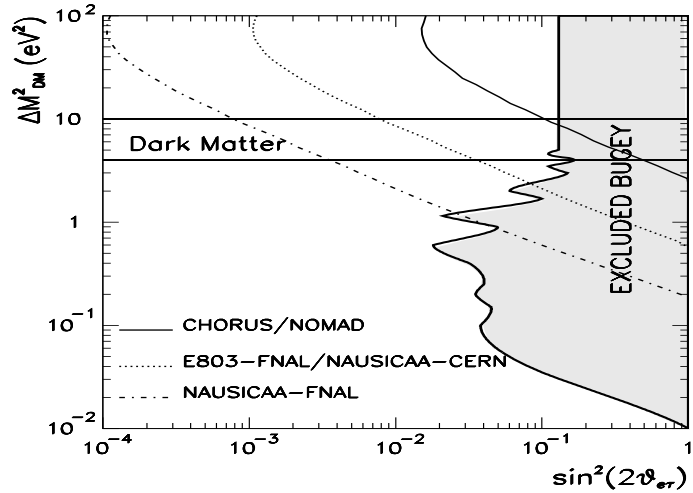,rwidth=2.5in,rheight=2.5in,width=4in,height=5in,bbllx=60pt,bblly=-30pt,bburx=500pt,bbury=750pt}
\caption{Accessible regions (90\% CL) for the $\nu_e\rightarrow \nu_\tau$
oscillation in the ($\Delta M^2_{DM}$, $\sin^2(2\theta_{e\tau})$) plane for the
CHORUS/NOMAD  experiments (fine solid line) and Fermilab experiment  E803/the
improved CERN experiment  (dotted line).  The dot-dashed line delimits the
region accessible to NAUSICAA at Fermilab.  Also shown in the figure are the
region at  present excluded  by the Bugey  data (shaded area) and the favoured
range on $\Delta M^2_{DM}$ from dark matter considerations (solid horizontal
lines).\label{etau}}
\end{figure}
\end{center}
The two upcoming $\nu_\mu(\nu_e) \rightarrow \nu_\tau$ experiments, CHORUS and
NOMAD \cite{CERN}, are $\nu_\tau$ appearance experiments, i.e., they search for
the
appearance of $\nu_\tau$'s  in  the CERN SPS beam consisting primarily of
$\nu_\mu$'s, with about 1\% $\nu_e$'s. The mean energy of the $\nu_\mu$ beam is
around $30$ GeV and the detectors are located approximately $800$ m away from
the beam source. Their expected
performance are summarized in Table \ref{tab:performance}.
There are a number of future $\nu_\mu(\nu_e) \rightarrow \nu_\tau$ experiments
being discussed at present.  As a specific example of
these experiments we have considered  a suggestion to upgrade the NOMAD
detector \cite{nomad01}. We will refer to this future detector with the generic
name of Neutrino ApparatUS with Improved CApAbilities (NAUSICAA). The detector
performance is summarized in Table \ref{tab:performance}.
\begin{figure}
\begin{center}
\psfig{figure=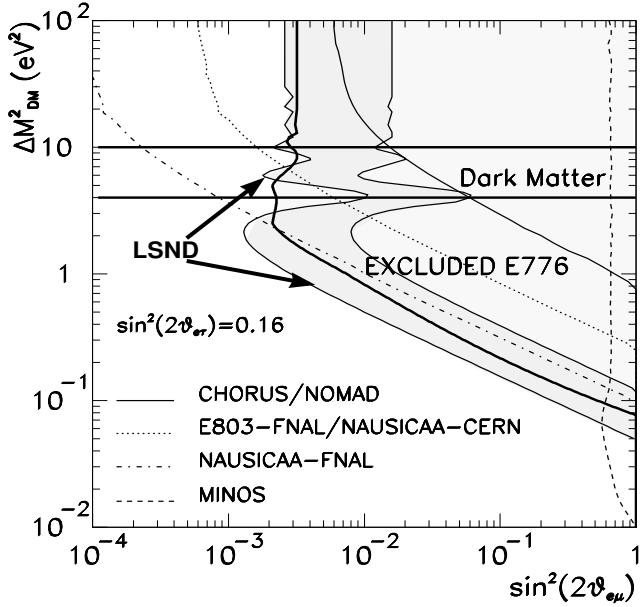,rwidth=2.5in,rheight=2.5in,width=4in,height=5in,bbllx=60pt,bblly=10pt,bburx=500pt,bbury=600pt}
\end{center}
\caption{Accessible regions (90\% CL) for the $\nu_\mu\rightarrow \nu_\tau$
optimum value of the mixing $e\tau$. The fine solid line corresponds  to the
CHORUS/NOMAD experiments and the dotted line corresponds to the Fermilab
experiment E803 and to the improved CERN experiment. The dot-dashed line
corresponds to NAUSICAA at Fermilab and the dashed line corresponds to MINOS.
Also shown in the figure are the regions at present excluded by E776  data
(shaded area) and the favoured range on $\Delta M^2_{DM}$ from dark matter
considerations (solid horizontal lines). For comparison also the  LSND data are
shown.\label{mutaudm}}
\end{figure}

At Fermilab, a neutrino beam will be available when the main injector becomes
operational, around the year 2001. Compared with the CERN SPS beam, the  main
injector will deliver a beam 50 times more intense, but with an average energy
around one third of that of the SPS neutrinos. There are currently two
experiments proposed to operate in this beam \cite{FNAL}. One is a
short-baseline experiment, E803, and  MINOS  a long-baseline experiment, which
proposes two detectors, separated by 732 km. This experiment can perform
several tests to look for a possible oscillation $\nu_\mu \rightarrow \nu_\tau$
in the small mass difference range. The expected
performance of E803 and MINOS are summarized in Table \ref{tab:performance}.
Finally we will consider the possibility of installing the NAUSICAA detector as
an alternative or a successor to E803 in the Fermilab beam which will improve
by one order of magnitude its sensitivity.

After implementing the limits derived in Sec.\ \ref{global} and considering the
sensitivity of the experiments, one sees that for all facilities the only
observable $\nu_e\rightarrow \nu_\tau$ transition oscillates with an
oscillation length $\Delta_{DM}$ such that $ P_{e\tau} \simeq
\sin^2(2\theta_{e\tau})\sin^2(\frac{\Delta_{DM}}{2})$. Figure \ref{etau} shows
the regions accessible to the experiments in  the
$(\sin^2(2\theta_{e\tau}),\Delta M^2_{DM})$ plane.   For transitions
$\nu_\mu\rightarrow \nu_\tau$ a four-neutrino framework predicts (unlike the
naive two-family framework) $two$ oscillations, dominated by the characteristic
lengths  $\Delta_{DM}$ and $\Delta_{AT}$. All experiments are in principle
sensitive to both oscillations depending on the values of the mixing angles
\begin{equation}  \begin{array}{lr} P^{DM}_{\mu\tau} \simeq
\sin^2(2\theta_{e\mu})\sin^2(\theta_{e\tau}) \sin^2(\frac{\Delta_{DM}}{2})
&\;\;\;P^{AT}_{\mu\tau} \simeq \sin^2(2\theta_{\mu\tau})\cos^2(\theta_{e\tau})
\sin^2(\frac{\Delta_{AT}}{2}) \\  \end{array} \label{mt}  \end{equation}  In
Fig.\ \ref{mutaudm} we show the regions accessible for the oscillation with
oscillation lenght $\Delta M^2_{DM})$ to the different experiments for an
optimum value of the $e\tau$ mixing.

\begin{figure}
\psfig{figure=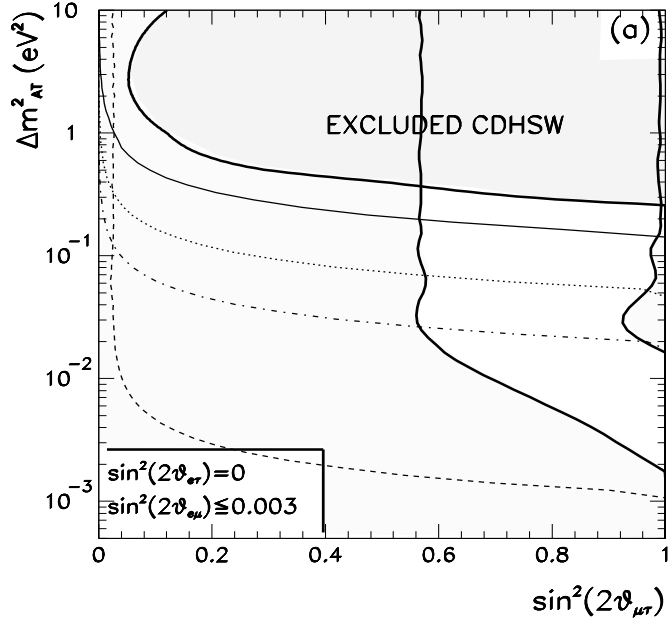,rwidth=2.5in,rheight=1.7in,width=4in,height=5in,bbllx=60pt,bblly=-60pt,bburx=500pt,bbury=530pt}
\caption{
Accessible regions (90\% CL) for the $\nu_\mu\rightarrow \nu_\tau$  oscillation
in the ($\Delta m^2_{AT}$, $\sin^2(2\theta_{\mu\tau})$) plane  for zero value
of the mixing $e\tau$. The fine solid line corresponds  to the CHORUS/NOMAD
experiments and the dotted line corresponds to the Fermilab experiment E803 and
to the improved CERN experiment. The dot-dashed line corresponds to  NAUSICAA
at Fermilab and the dashed line to MINOS. Also shown in the figure are the
regions at present excluded  by CDHSW data  (dark shaded area) and the
atmospheric neutrino analysis (light shaded area).\label{mutauat1}}
\end{figure}

\begin{figure}
\psfig{figure=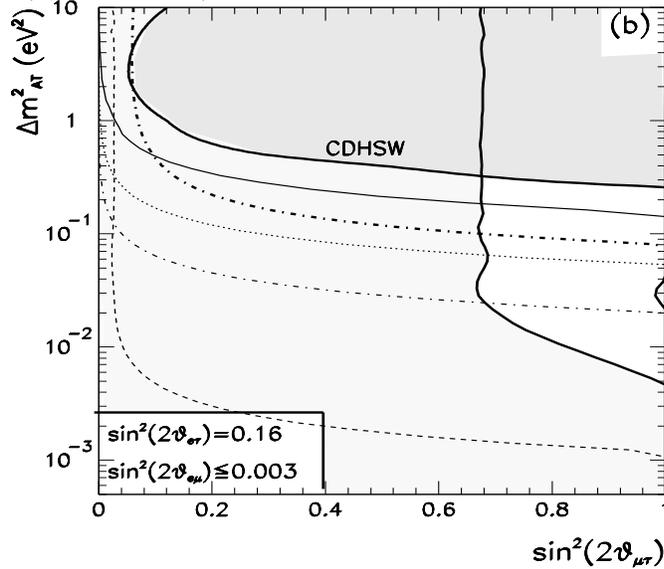,rwidth=2.5in,rheight=2.5in,width=4in,height=5in,bbllx=60pt,bblly=10pt,bburx=500pt,bbury=650pt}
\caption{
Same as previous figure for maximum mixing $e\tau$.  Also shown in the
figure are the region at present excluded  by CDHSW  data (dark shaded area)
and the atmospheric neutrino analysis (light shaded area). These are the only
constraints for  zero or values of the $e\mu$ mixing $\sin^2(2\theta_{e\mu})\ll
0.003$. For maximum value  $\sin^2(2\theta_{e\mu})\simeq 0.003$ the E776 limit
(thick dot-dashed line) is relevant.\label{mutauat2}}
\end{figure}

Figures \ref{mutauat1} and \ref{mutauat2} show the region accessible to
the experiments for the oscillation in $\Delta m^2_{AT})$ for different
values of the other mixings.

\end{document}